# Pressure dependence of direct optical transitions in ReS$_2$ and ReSe$_2$

R. Oliva,[1†*] M. Laurien,[2†] F. Dybala,[1] J. Kopaczek,[1] Y. Quin,[3] S. Tongay,[3] O. Rubel,[2] and R. Kudrawiec[1]

[1]Department of Experimental Physics, Faculty of Fundamental Problems of Technology, Wroclaw University of Science and Technology, Wybrzeże Wyspiańskiego 27, 50-370 Wrocław, Poland
[2]Department of Materials Science and Engineering, McMaster University, JHE 359, 1280 Main Street West, Hamilton, Ontario L8S 4L8, Canada
[3]Department of Materials Science and Engineering, University of California, Berkeley, California 94720, USA
* Corresponding author: robert.oliva.vidal@pwr.edu.pl
† These authors contributed equally to this work

## Abstract

We present an experimental and theoretical study of the electronic band structure of ReS$_2$ and ReSe$_2$ at high hydrostatic pressures. The experiments are performed by photoreflectance spectroscopy and are analyzed in terms of *ab initio* calculations within the density functional theory. Experimental pressure coefficients for the two most dominant excitonic transitions are obtained and compared with those predicted by the calculations. We assign the transitions to the Z k-point of the Brillouin zone and other k-points located away from high-symmetry points. The origin of the pressure coefficients of the measured direct transitions is discussed in terms of orbital analysis of the electronic structure and van der Waals interlayer interaction. The anisotropic optical properties are studied at high pressure by means of polarization-resolved photoreflectance measurements.

## I. INTRODUCTION

The ReX$_2$ crystals (X = S, Se) are semiconductors from the family of two-dimensional layered transition metal dichalcogenides (TMDCs) that exhibit special properties. Rhenium-based TMDCs have received increasing interest during the last few years owing to their large in-plane anisotropic properties. These properties result from their particular band structure and reduced crystal symmetry (i.e. $T_d$ distorted octahedral structure, belonging to the triclinic crystal system[1]) as well as their strong 2D character that has been attributed to weak van der Waals interlayer bonding even in their bulk form.[2,3] Besides the large fundamental interest, ReX$_2$ has also shown to be a highly interesting technological material for many potential applications, including photodetectors,[4–9] solar cells,[10] photonics,[11] flexible electronics[12] and field-effect transistors.[13–17] Remarkably, the small interlayer coupling of ReX$_2$ opens an exciting field of new possibilities, as it may allow to design bulk devices that retain 2D functionalities only present in single layered materials.[18] In order to fully exploit the applications of ReX$_2$ for developing novel optoelectronic devices, it is crucial to further characterize its fundamental properties.

Optical modulation spectroscopy is a very powerful method to study the optical properties of semiconductors. Owing to its differential-like character, interband related features are highly enhanced and background signal is suppressed, thus allowing to accurately measure direct optical transitions.[19] So far, different modulation spectroscopies have shown to be very useful for studying the optical transitions of ReX$_2$: Early temperature-dependent piezoreflectance measurements revealed that at ambient temperature ReSe$_2$ and ReS$_2$ exhibit one and two strong excitonic transitions, respectively.[20] Furthermore, transmission and thermoreflectance measurements showed that the polarization dependence of the amplitude of these transitions is described by a sinusoidal function, evidencing the pronounced anisotropic properties of ReX$_2$.[21] Authors in another were able to tentatively assign the interband origin of such excitons by comparing first-principle calculations with electrolyte electroreflectance measurements.[22] Later works performed polarization-dependent experiments with different optical techniques and identified a total of three and four excitonic transitions for ReS$_2$ and ReSe$_2$, respectively.[23,24,25] These works provided evidence that these excitons, which exhibit strongly polarized dipole character, were confined within single layers. This is in agreement with recent works that show excitons to be confined within a single layer by 68%.[24]

However, the extent to which ReX$_2$ behaves as stacked decoupled layers has recently been a topic of intense discussion.[3,23,24,26–28] On the one hand, recent photoemission spectroscopy measurements reveal a significant electronic dispersion along the van der Waals gap, thus confirming a certain degree of electronic coupling.[23,26,27] Also, recent calculations show that the fundamental bandgap shrinks by 32.7% from monolayer to bulk and the interlayer binding energy is similar to other TMDCs such as MoS$_2$.[29] On the other hand, optical, vibrational and structural measurements indicate that ReX$_2$ exhibits a strong 2D character. For instance, PL measurements revealed that, in contrast with other TMDCs, emission energy barely shifts between monolayers and bulk ($\Delta E \approx -50$ meV for ReS$_2$ while $\Delta E \approx -600$ meV for MoS$_2$), and the intensity increased with increasing the number of layers, rather than vanishing from a direct-to-indirect transition typically observed in conventional TDMCs.[3] One of the most direct ways to probe interlayer interaction is by performing high-pressure measurements which allow to directly modulate the interlayer distance. In this regard, high-pressure XRD measurements revealed that ReX$_2$ exhibits the largest compressibility amongst all TMDCs,[30,31] and high-pressure Raman measurements on ReS$_2$ showed a twofold decreased pressure coefficient of the out-of-plane $A_{1g}$ phonon mode with respect to other TMDCs.[3] spite the fundamental properties of this crystal system being relatively well known at ambient pressure, high-pressure optical measurements would provide highly valuable information in order to evaluate the degree of electronic interlayer coupling in ReX$_2$.

High-pressure optical measurements are widely employed to obtain detailed structural and band structure information of semiconductors.[32] Moreover, high-pressure optical measurements provide a highly useful benchmark to test first principle calculations (such as those based on density functional theory) on challenging systems such as TMDCs. For the case of ReX$_2$, which exhibit weak interlayer forces at ambient pressure, high-pressure optical measurements would shed new light into the role of orbital composition and van der Waals bonding on the excitonic energies and their pressure dependency. To date, the amount of high pressure optical studies on ReX$_2$ is scarce. The pressure dependency of the band gap has only been experimentally investigated for ReS$_2$ by means of photoluminescence and absorption.[33,3] These works found that the band gap of ReS$_2$ does not increase with pressure and an almost-direct to indirect band gap transition takes place around 27 kbar. At higher pressures calculations suggest that ReS$_2$ exhibits a metallization and superconducting state.[34]

Despite the previous investigations, there are still many questions that remain to be addressed with regard to the high-pressure optical properties of ReX$_2$. Firstly, an experimental assignment of the different excitonic transitions around the band gap is desirable. So far piezoreflectance measurements on ReSe$_{2-x}$S$_x$ alloy suggested that the nature of the direct band edges are similar for each compositional end member,[35] but electronic dispersion calculations together with angle-resolved photoemission spectroscopy (ARPES) measurements suggested that the first direct electronic transitions take place either at the Z high-symmetry point of the Brillouin zone (BZ) or away from the zone center, far from any particular high symmetry direction.[26,27,36,37,29] Secondly, while the orbital composition of the states of ReS$_2$ has been described for different numbers of layers,[29] the interplay of orbital composition on the pressure dependence on the electronic band structure has not yet been investigated. Finally, the anisotropic properties of ReX$_2$ at high pressure still remain to be explored.

Here we report photoreflectance (PR) measurements performed at high hydrostatic pressure on thin ReS$_2$ and ReSe$_2$ exfoliated flakes. In order to ensure the reproducibility of the experimental results, samples obtained from different sources and grown in different conditions are used for the experiments. Polarization-dependent measurements performed at different pressures are used to energetically resolve the different excitonic transitions that exhibit very similar energies. Our results show that the two main direct transitions for ReS$_2$ and ReSe$_2$ exhibit a negative pressure coefficient, in contrast to other TMDCs such as MoS$_2$, MoSe$_2$, WS$_2$ or WSe$_2$.[38] Such finding provides valuable information to assess the degree of electronic interlayer coupling and role of orbital composition on the energies of the band edge states. The experimental results are discussed in the light of *ab initio* band structure calculations. These calculations are performed using different functionals and considering different hydrostatic pressures. Good agreement is found between the experimental and calculated pressure coefficients for the two main transitions. By inspecting the calculated electronic dispersion curves along a large grid of k-points in the whole 3D Brillouin zone we are able to assign the experimentally observed transitions. Finally, the negative sign of the measured pressure coefficients are discussed in terms of orbital contributions to the states of the valence and conduction band of each transition and van der Waals interaction.

## II. EXPERIMENT

For the present work two samples of different origins were used for each set of $ReS_2$ and $ReSe_2$ materials. One sample for each material was commercially obtained from HQgraphene, which consisted of thin flakes mechanically exfoliated from synthetic bulk crystals (99.995 % purity). These are here labelled as samples I and III for $ReS_2$ and $ReSe_2$, respectively. The $ReS_2$ (sample II) and $ReSe_2$ (sample IV) samples were synthesized by the chemical vapor transport growth technique using Re (99.9999% purity), S or Se (99.9999% purity) pieces. These precursors were mixed at atomic stoichiometric ratios and sealed into 0.5 inches diameter and 9 inches long quartz tubes under $10^{-6}$ Torr. Extra ReI3 was added as a transport agent to initiate the crystal growth and successfully transport Re, S, and Se atomic species. Closely following Re-S-Se binary phase diagrams, we have synthesized crystals with temperature variation (drop) of 50 °C over 5 weeks to complete the growth. Samples were cooled down to room temperature and ampoules were opened in a chemical glove box. The use of two samples grown under different conditions for each material allow to further validate the reproducibility of the here presented experimental results.

In order to perform the high-pressure hydrostatic measurements, the samples were mounted inside a UNIPRESS piston cylinder cell. The chosen pressure hydrostatic medium was Daphne 7474 which remained hydrostatic and transparent during the whole measurements, up to pressures of 18 kbar. The pressure was determined by measuring the resistivity of a InSb probe which provides a 0.1 kbar sensitivity. A sapphire window in the press allowed optical access to perform photoreflectance measurements (PR). For the PR measurements a single grating of 0.55 m focal length and a Si pin diode were used to disperse and detect the light reflected from the samples. A chopped (270 Hz) 405 nm laser line was pumped into the sample together with a probe tungsten lamp (power of 150 W). Phase-sensitivity detection of the PR signal was processed with a lock-in amplifier. Further details on the experimental setup can be found elsewhere.[39] All measurements were performed at ambient temperature and pressures up to ≈18 kbar. At this pressure range no phase transition was observed and only the $T_d$ crystal structure was investigated.

# III. *AB INITIO* CALCULATIONS

## A. Computational details

*Ab initio* calculations on the density functional theory (DFT) level were carried out using the Vienna Ab initio Simulation Package (VASP)[40,41] with the projector augmented wave[42] (PAW) potentials as implemented by Kresse and Joubert.[43] The strongly constrained and appropriately normed (SCAN)[44] semilocal exchange-correlation functional was employed. SCAN belongs to the meta-general-gradient-approximation (meta-GGA) functionals and has shown to produce more accurate results than conventional GGA functionals at a very comparable computational cost.[44–46] In particular, SCAN is recommended for electronic structure prediction of materials with heterogeneous bond types[45] (e.g. covalent and van der Waals) as well as layered materials.[46] It is therefore well suited for the band structure prediction of $ReX_2$. In addition, a revised Vydrov-van Voorhis (rVV10) long-range van der Waals interaction[47–49] was used.

Structure information of $ReS_2$ and $ReSe_2$ was taken from Murray et al.[50] and Alcock and Kjekshus,[51] respectively. The original lattices were transformed respective to their coordinate system to give credit to the convention of defining the Cartesian *z*-axis as perpendicular to the layers and lattice vector *c* as the out-of-plane vector crossing the 2D-layers (*Figure 1 b, d*). It should be noted that due to the offset in layer stacking the *c*-vector is not perpendicular to the layers. $ReX_2$ crystallizes in the $T_d$ distorted octahedral configuration. In this configuration four Rhenium atoms group together to form so-called diamond-like clusters that are assembled in chains within the plane (*Figure 1 b, d*, right side, diamond chains are highlighted in red). The origin of the distortion in the crystallographic structure has been discussed in the light of the Jahn-Teller effect as well as quasi 1D-Peierls distortion.[52–54]

Structure relaxation was undertaken with a Monkhorst-Pack[55] k-mesh of $5 \times 5 \times 5$ with the above-mentioned basis set and functionals. Seven electrons were considered for the valency of Re ($5d^5 6s^2$). The cutoff energy for the plane-wave expansion was set to 323.4 eV and 282.8 eV for $ReS_2$ and $ReSe_2$, respectively, which is 25 % above the recommended values in the pseudopotential files. The desired properties (pressure coefficient, band gaps and band character) were carefully checked for convergence with the kinetic energy cutoff as it can be seen in the Figures S9-S14 of the Supplementary Information, (S.I.). Good convergence is reached at 550 eV for $ReS_2$ and 400 eV for $ReSe_2$, however the above mentioned kinetic cutoff values are sufficient for our purposes.

Fully relaxed structures were calculated for pressures of 0, 5, 10, 15 and 20 kbar. Pressure was applied by adding desired pressure values to the stress tensor (*PSTRESS*-tag). The structures were relaxed until the total energy change and the band structure energy change dropped below $10^{-7}$ eV and the residual atomic forces were less than 0.02 eV/Å in their absolute value. Crystallographic information files (CIF) with atomic structures at 0 and 20 kbar as used in the calculations can be accessed through the Cambridge crystallographic data center (CCDC deposition numbers 1862132-1862135).

## B. Band structure calculations

For calculations of the band structure and optical properties, spin-orbit interaction was taken into account. The cutoff energy was set to normal accuracy which is 258.7 eV for $ReS_2$ and 226.2 eV for $ReSe_2$. High-density Gamma-centered k-mesh calculations ($34 \times 34 \times 34$) were performed to investigate possible valence band maxima (VBM) and conduction band minima (CBM) located off the symmetry points. In addition, the k-path was prepared based on the suggestions of the seeK-path tool[56], according to which the path walks through the first Brillouin zone whilst returning frequently to the Γ point in order to intersect the space in a three-dimensional manner (see *Figure 1 c* for illustration). Scaled reciprocal coordinates are listed for the special k-points in Table 1. The coordinates of the high-symmetry k-points can be found in the S.I. (Table S1).

*Table 1.* Scaled reciprocal coordinates of special *k*-points of ReX$_2$ corresponding to the global valence band maximum (VBM) and conduction band minimum (CBM), as well as the k-points of the assigned direct transitions A and B.

| Material | k-point | a* | b* | c* | Function |
|---|---|---|---|---|---|
| ReS$_2$ | Z | 0.0 | 0.0 | 0.5 | Transition A & CBM |
|  | K1 | 0.05882 | −0.05882 | 0.38235 | Transition B |
|  | K2 | 0.20588 | −0.17600 | 0.294118 | VBM |
| ReSe$_2$ | J1 | 0.02941 | 0.14706 | −0.20588 | Transition A |
|  | Z | 0.0 | 0.0 | 0.5 | Transition B |
|  | J3 | 0.20588 | 0.38235 | 0.20588 | CBM |
|  | J2 | 0.08824 | 0.14706 | −0.20588 | VBM |

Besides the energy eigenvalues, the optical matrix elements $M_{nm,\alpha}(k) = \langle \varphi_n(k) | d/dk_\alpha | \varphi_m(k) \rangle$ between bands $n$ and $m$ were calculated within a linear optical theory. The matrix element consists of three complex numbers. To estimate the scattering efficiency of the optical transition we evaluated the squared sum of the absolute values, $\sum_{\alpha=1}^{3} |M_{nm,\alpha}|^2$.

On a technical note, the number of bands was increased to 320 bands for the optical matrix element calculation, which is roughly a 2.5-fold value of the VASP default. This is necessary because the matrix calculation requires a large number of empty bands. Also, the number of frequency grid points (VASP: *NEDOS*-tag) was increased to 6000.

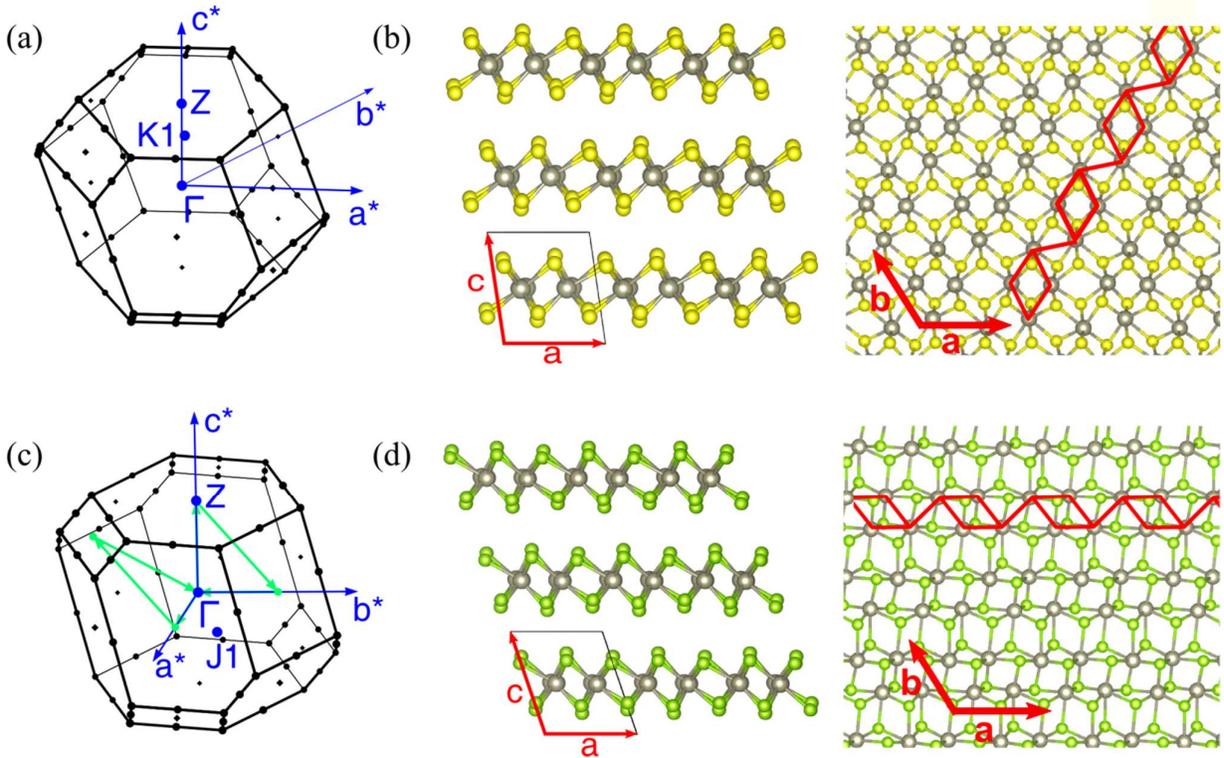

*Fig. 1.* Structural information of ReX$_2$. Brillouin zone of ReS$_2$, (a) and ReSe$_2$ (c). Atomic configurations of the distorted $T_d$ structure for ReS$_2$ (b) and ReSe$_2$ (d).

## IV. RESULTS

### A. ReS$_2$

#### a. High-pressure measurements

The photoreflectance spectra obtained for ReS$_2$ at different pressure values are shown in Fig. 2. As it can be seen in the figure, two main features can be observed for either sample I or II (upper and lower panel, respectively) which correspond to the direct transitions A and B. The intensity of the transition A is significantly stronger than that of transition B, and both energies decrease with increasing pressure. The evolution of energies with increasing pressure is qualitatively shown in the figure as two straight lines for each transition. At ≈15 kbar both features merge, since transition B exhibits a larger redshift with increasing pressures. Previous polarization-resolved modulated spectroscopic experiments at ambient pressure showed that ReS$_2$ exhibits two strong excitonic transitions at ambient temperature[20,21] and a weaker transition with slightly higher energy,[22] in good agreement with our measurements at $P = 0$ kbar. The third transition is resolved with polarization measurements, as shown in the next section, also at different pressures.

In order to determine the energy of the transitions from the PR spectra, these were fitted with the Aspnes formula,[57] given by

$$\frac{\Delta R}{R}(E) = \text{Re}\left[\sum_{j=1}^{n} C_j e^{i\theta_j}(E - E_j + i\Gamma_j)^{-m}\right], \qquad [1]$$

where $n$, $C$ and $\theta$ are the number of transitions, amplitude and phase of the resonance, $E_j$ and $\Gamma$ are the energy and broadening parameter of the transition, respectively. For excitonic transitions we take $m = 2$. Two transitions are enough to successfully reproduce all the spectra shown in Fig. *2* (dotted curves). For the fitting procedure we left all parameters unfixed for the spectrum obtained at ambient pressure, while only the amplitude and the energy of the transition were left as free parameters for spectra at higher pressures, since these are expected to change with pressure. The pressure dependence of the energy of each transition is plotted in Fig. *3*. for both samples.

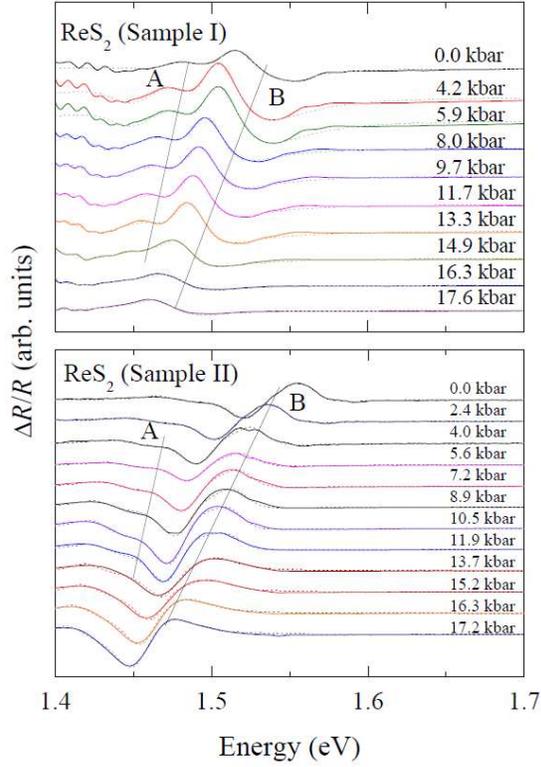

*Fig. 2.* Photoreflectance spectra obtained at different pressures for sample I (top panel) and sample II (bottom panel). Fittings are shown as dotted curves. Straight lines around the fitted transition energies are shown as guide to the eye for transitions A and B. The spectra at low pressure values exhibit one strong transition (i.e. transition B) and a weaker one (transition A). Both features decrease in energy with increasing pressure and merge at high pressures.

As it can be seen in Fig. *3*, the energy of the transition A (shown as squares) decreases with increasing pressure at a lower rate than the energy of the transition B (shown as circles). No data points are shown for the transition A at pressures below 4 kbar because at low pressures the amplitude of this transition is too weak and exhibits features of a third transition, which lead to inaccuracies in the fitted energies. The differences in energies for the B transition between both samples are accounted by both, weak signal and sample misorientations inside the pressure cell. The relative orientation of each sample, and contributions from a third transition are investigated in the next section, where polarization-dependent measurements are presented at selected pressures. Here, a linear fit was performed for the data of both samples combined to establish a measure of the pressure dependency of the transition energies. This yields a pressure coefficient of −2.3 meV/kbar and −4.2 meV/kbar for the transitions A and B, respectively. So far, the amount of experimental works dealing with the high-pressure optical properties of ReS$_2$ is very scarce. High-pressure photoluminescence measurements showed that the main emission peak of ReS$_2$ exhibits a redshift of around −2.0 meV/kbar,[33] which we attribute to the A transition. On the other hand, high-pressure absorption measurements showed that the absorption edge is rather insensitive to pressures up to 7 GPa,[3] but a redshift with increasing pressure takes place in the absorption edge around the optical gap energy (i.e. $E_g \approx 1.55$ eV). At higher pressures, the band gap closes and a metallization has been reported at 354 kbar.[58]

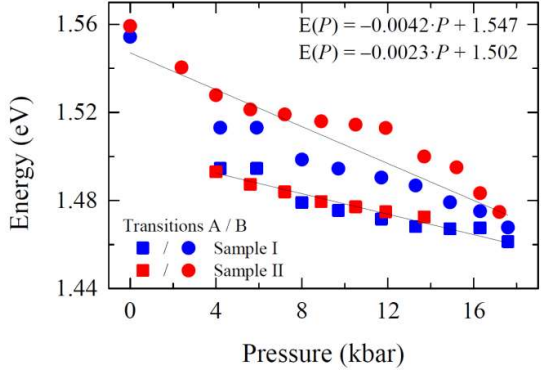

*Fig. 3.* The energy of the fitted transitions in the photoreflectance experiments is plotted as a function of pressure for samples I and II in blue and red symbols, respectively. The stronger transition, at higher energy (transition B, shown as circles), exhibits a pressure coefficient of −4.2 meV/kbar, while the lower energy transition (transition A, shown as squares) exhibits a lower pressure coefficient, around −2.3 meV/kbar.

**b. Polarization measurements**

Since the broadening parameter of the excitonic transitions at ambient pressure is around ≈ 20 meV, it is difficult to spectrally resolve and fit all the transitions that coexist in a range smaller than ≈ 40 meV with Eq. [1]. In order to evaluate all the excitonic transitions and their pressure dependence, polarization-resolved measurements have been performed at selected pressures (all spectra are shown in the S.I., Figs. S1 and S2). From these polarization-dependent spectra the Aspnes formula was fitted assuming that the phase, energy and broadening parameters are independent of polarization angle, and taking only the amplitude as a free parameter. Within this method, we found that three transitions are enough to describe all the features present in the PR spectra at all polarization angles. The energies of the three fitted transitions were within a spectral range of ≈60 meV. These results are in agreement with previous polarization-resolved measurements, which also report three close excitonic transitions within a spectral range of ≈100 meV.[22,23,59] The angular dependence of the amplitude is plotted in Fig. 4 for the three excitonic transitions, namely $E^{ex}_1$, $E^{ex}_2$, and $E^{ex}_3$.

The polarization dependence of the amplitude of the fitted excitonic transitions is shown for pressures of 10 kbar (open circles) and 16.3 kbar (full circles) in Fig. 4. and are plotted with a phase separation of 180 degrees to facilitate the comparison between pressures. By comparing the upper and lower half of the chart, it is clear that the amplitude and orientations for each transition is maintained with increasing pressure. This indicates that the sample did not suffer any structural transition, which is expected since the first phase transition takes place at higher pressures, around 113 kbar.[31] The angular dependence of the amplitudes is fitted by using a formula derived from the Malus law,

$$f(\varphi) = f_\parallel \cos^2(\varphi - \varphi_0) + f_\perp \sin^2(\varphi - \varphi_0), \qquad [2]$$

where $f_\parallel$ and $f_\perp$ are the parallel and perpendicular components of the oscillator strength, and $\varphi_0$ the relative orientation of the excitonic transition. Following Fig. 4, the angular dependence of the amplitude of the most energetic transition (i.e. $E^{ex}_3$) is very weak, in agreement with previous polarization-resolved piezoreflectance measurements.[59] On the other hand, the first and second excitonic transitions (i.e. $E^{ex}_1$ and $E^{ex}_2$) are strongly polarized ($f_\perp \approx 0$) along the angles $\varphi_0 \approx 4°$ and 93°, respectively. These results are in relatively good agreement with previous works focused on the anisotropic properties of bulk ReS$_2$ by means of polarization-resolved PL performed at a temperature of 110 K.[60] This work found that the orientation of the $E^{ex}_1$ and $E^{ex}_2$ relative to the *b*-axis is $\varphi_0 \approx 17°$ and ≈ 86°, respectively. During the loading process our samples were misoriented inside the press cell, but from the angular dependence of the excitons it was possible to determine the orientation of the *b*-axis for each sample with respect the vertical axis. These are $\varphi \approx 97°$ and −13° for samples I and II, respectively.

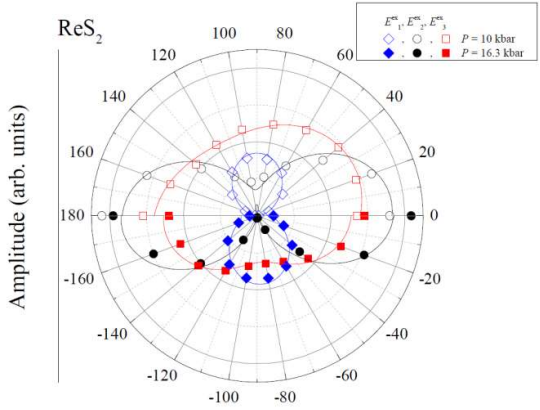

*Fig. 4.* Polar dependence of the amplitude for three fitted PR transitions in Sample II. The fittings have been performed at two different pressure values, 10.0 kbar (empty symbols, top) and 16.3 kbar (filled symbols, bottom). The solid curves are fits from Eq. [2].

### c. First-principle calculations

First principle calculations were carried out in order to assign the experimentally observed transitions to locations within the Brillouin zone and to provide further insight into the electronic and optical properties of ReS$_2$. After structure relaxation at a pressure of 0 kbar, the calculated (experimental[61]) lattice parameters are $a$ = 6.438 Å (6.450 Å), $b$ = 6.498 Å (6.390 Å), $c$ = 6.357 Å (6.403 Å) and $\alpha$ = 106.556° (105.49°), $\beta$ = 88.154° (91.32°), $\gamma$ = 121.359° (119.03°) for ReS$_2$. With increasing pressure, the $c$ lattice parameter decreases most strongly, reaching 97.7 % of its initial value for pressures of 20 kbar. This is expected considering that the lattice vector $c$ points across the individual layers of ReS$_2$ that are held together by weak van der Waals forces only. Lattice parameters $a$ and $b$ decrease by less than 0.5 % and the lattice angles vary only slightly (see S.I., Section 3, for more details).

High-density k-mesh calculations were conducted with a k-mesh of 34 × 34 × 34 to ensure a complete investigation of the 3D Brillouin zone for valence band and conduction band extrema. At 0 kbar pressure, our calculations predict a quasi-direct fundamental band gap at the Z-point of 1.20 eV. The global valence band maximum is actually found at K2, being only 2 meV higher in energy than at the Z point. This occurs due to the many close-lying valence band maxima in ReS$_2$. We also performed calculations at a higher level of theory using the hybrid functional HSE06[62] for comparing the Z and K1 point. At this level of theory, the global valence band maximum is located at the Z point. This allows us to conclude for a quasi-direct (or marginally indirect) band gap. Further, we found a second direct transition off the high symmetry points, at K1 with a band gap energy of 1.21 eV. K1 is located roughly between Γ and Z (see Table 1). The calculated gaps of around 1.20 and 1.21 eV are smaller than the measured optical gaps of around 1.50 and 1.55 eV. This is expected as the SCAN functional typically exhibits a band gap underestimation of 15% for transition metal dichalcogenides.[46] Our HSE06 Calculations were able to predict the band gap energies more accurately with 1.42 eV and 1.43 eV at Z and K1 respectively after subtracting the excitonic binding energy (see Section 6 of the S.I.). The location of the fundamental gap of ReS$_2$, predicted by our calculations to be at Z, is in agreement with recent direct measurements of the band dispersion using ARPES[36,26] and a recent theoretical study employing quasiparticle approximations.[29]

Owing to the complex band structure of ReS$_2$, which exhibits very close direct and indirect band gaps in energy and position in k-space,[26,29] some computational considerations should be taken into account. For instance, some theoretical works predict the valence band maximum of bulk ReS$_2$ to be either at Γ[3,33,63] or at Z[64] but with deviations from the experimental results. This is a consequence of choosing only a few high-symmetry paths for the calculation of the band structure, or functionals that fail to capture the details of the complex electronic structure of ReX$_2$. Hence, it is clear that in order to accurately describe the band structure of complex materials like ReX$_2$, it is important to consider the whole 3D-Brillouin zone, and to devote careful attention to the choice of the functional. Recent contributions show that the choice of the functional influences the number and location of VBM and CBM in ReX$_2$.[26,37,64] The results of the meta-GGA functional SCAN employed here seem to reproduce experimental results well and at a low computational cost.

At pressures of 20 kbar the high-density k-mesh calculations predict an indirect fundamental band gap of 1.15 eV for ReS$_2$ with its VBM located at K2 and its CBM at Z (see Table 1). The two lowest-energy direct band gaps stay at Z and K1 with increasing pressure, with a more pronounced band gap reduction for the K1 transition. The narrowing of the band gap observed is in general agreement with other findings evaluating the pressure dependence of bulk ReS$_2$.[34,58]

The electronic band structure and the optical matrix element were calculated for *k*-path intersecting the first Brillouin zone in a 3-dimensional manner. *Figure 5 (A)* shows electronic dispersion curves for 0 kbar (black lines) and 20 kbar (red lines) calculated with the SCAN functional. The normalized values of the optical matrix element as well as the variation of the band gap along the path are also shown. Clearly seen are the two primary direct transitions at Z and K1 with band gap energies of 1.20 eV and 1.21 eV, respectively. The matrix element reaches high values around these k-vectors, which confirms the optical activity of these transitions.

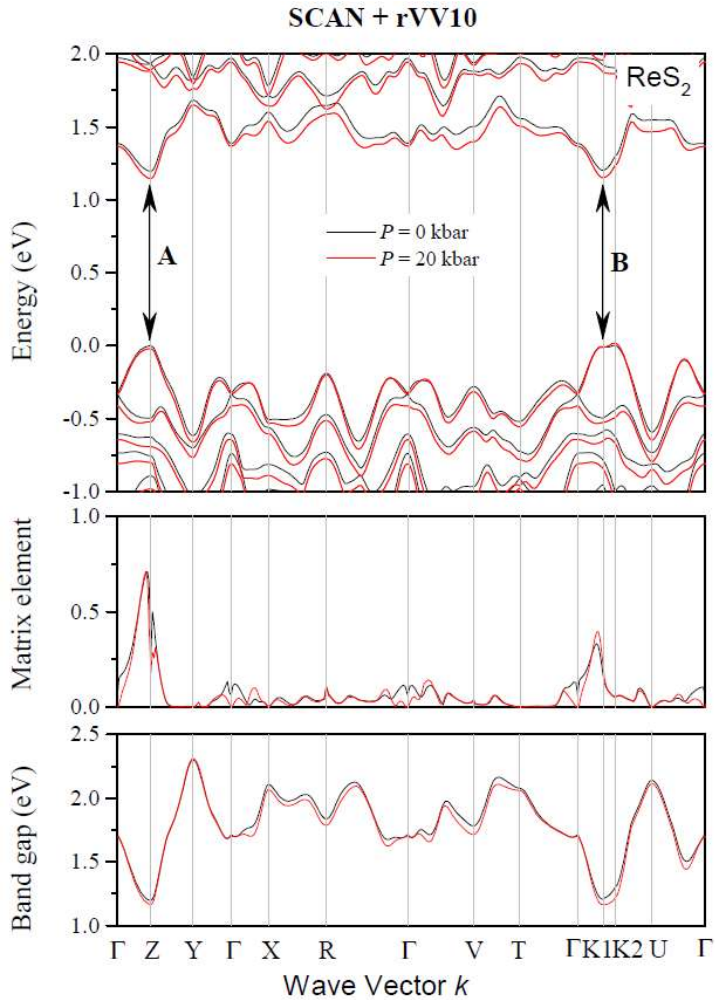

Fig. 5.

Electronic dispersion curves for ReS$_2$ as calculated using the SCAN functional at zero pressure (black curves) and 20 kbar (red curves). The corresponding matrix elements have been calculated for each *k*-point, the stronger transitions are located around the Z and K1 *k*-points. In the lower panels, the direct band gap energy is plotted along the studied *k*-path.

In order to assign the experimentally determined transitions, the pressure dependence of the calculated band gap energies at K1 and Z are compared with the A and B excitonic energies for samples I and II measured by photoreflectance. This can be seen in Fig. 6, where the pressure dependency of the variation of energy is plotted for both, calculated band gaps and measured excitonic transitions. The figure, which plots variation of energy rather than absolute values, allows to directly compare theoretical calculations (open symbols) with experimental results (full symbols) neglecting energetic differences arising from the DFT band gap underestimation and excitonic binding energy. The calculated pressure coefficient is taken from a linear fit to bandgap energies calculated at different pressures (see Section 4 of the S.I. for more details); we obtained values of of −1 meV/kbar$^{-1}$ and −3.4 meV/kbar$^{-1}$ for Z and K1, respectively. As it can be seen in the figure, our calculations predict distinct pressure coefficients for each transition, which is also observed experimentally. While the calculated pressure coefficients differ from the measured ones, around −2.3 meV/kbar and −4.2 meV/kbar for transitions A and B respectively, they fall within the experimental and calculation uncertainties. Note that the calculated energies are somewhat scattered, which yields to a statistical error of the pressure coefficient around ±0.6 meV/kbar for Z and ±1.2 meV/kbar for K1. This is due to the method employed: the band energies were obtained by taking the k-points that corresponded to the maximum of the matrix elements around the Z and K1 k-points, which shifted slightly with increasing pressure mostly due to structural distortions. Despite the uncertainties associated with this method they provide a realistic description of the pressure dependency of the band gaps since the matrix element is proportional to the light absorption. After comparing the pressure coefficients, transition A is assigned to the Z k-point and transition B to the K1 k-point.

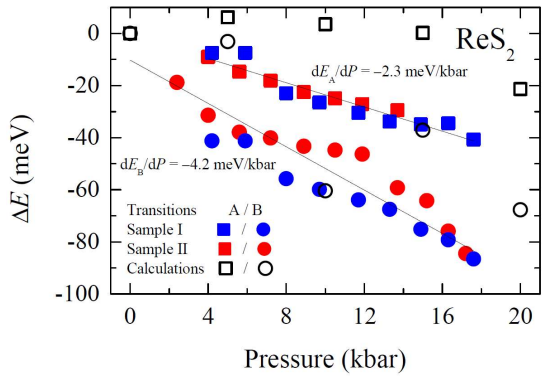

*Fig. 6.*
Increment of exciton energy versus pressure plotted for the transitions A (squares) and B (circles) from measurements on samples I (blue) and II (red) as well as the calculated values of the transport band gap using the SCAN functionals (open symbols). The straight lines are linear fits, the pressure coefficients of each transition are included.

## B. ReSe$_2$
### a. High-pressure measurements

Photoreflectance spectra for ReSe$_2$ at different pressures are shown in Fig. 7 for samples III (upper panel) and IV (lower panel). Two transitions can be identified in the spectra, a strong one (transition A) and a weaker at higher energy (transition B). This result is in agreement with previous piezoreflectance and optical absorption measurements at ambient conditions, which found two excitonic transitions.[65] However, it is worth noting that other polarization-dependent optical measurements performed at low temperature (≈15 K) found a higher number of excitonic transitions (at least a total of four) within a spectral range of 100 meV.[24,25] These transitions can be tracked only using polarization measurements and at low temperature since the broadening energy of the transitions at room temperature is too large, around 20-50 meV. As it can be seen in the figure, at ambient pressure the PR spectra resembles that of a single transition, and only two excitonic transitions can be clearly distinguished at higher pressures. Such splitting is a consequence of different pressure coefficients for both transitions. The energy of both transitions have been determined fitting Eq. [1] employing the same procedure used for ReS$_2$.

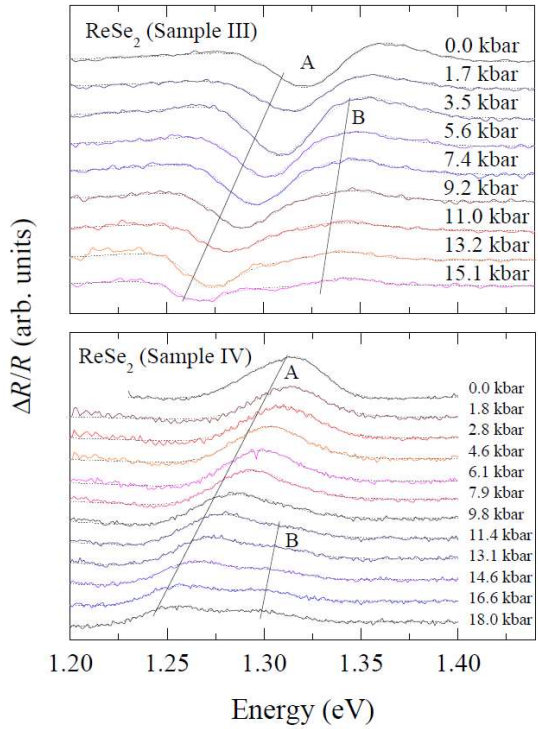

*Fig. 7.*
Photoreflectance spectra obtained at different pressures for sample III (top panel) and sample IV (bottom panel). Fittings are shown as dotted curves. Straight lines around the fitted transition energies are shown as guide to the eye for transitions A and B. The spectra at low pressure values exhibit one clear transition (i.e. transition A), and a second, weaker transition can be distinguished at higher pressures (transition B).

Fig. 8 shows the energy of the PR transitions fitted for samples III and IV as a function of pressure (blue and red symbols, respectively). It can be seen that the pressure dependency of the energy of transition A (squares) is nearly identical for both samples. This is not the case for the transition B, which was not possible to be fitted at pressures lower than 12 kbar for sample III and exhibits a shifted energy between both samples. We attribute the shifted energies to uncertainties in the fitting procedure, since both transitions are energetically very close to each other (around 60 meV) and transition B is significantly weaker than the main transition, as well as sample misorientations. Despite the uncertainty on the transition energy at low pressures for the transition B, pressure coefficients were obtained from linear fittings. These are −3.5 meV/kbar and −1.3 meV/kbar for the transitions A and B, respectively.

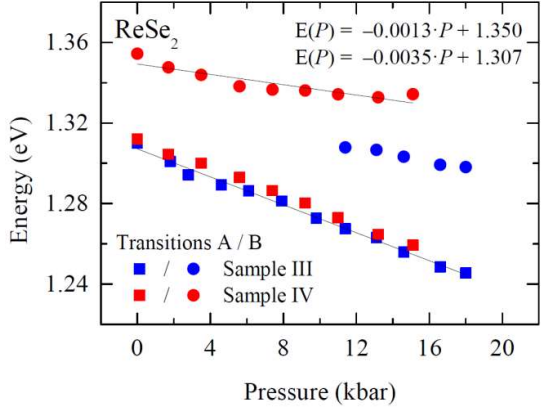

*Fig. 8.* The energy of the fitted transitions in the photoreflectance experiments is plotted as a function of pressure for samples III (blue symbols) and IV (red symbols). The main transition (transition A, shown as squares) exhibits a pressure coefficient of −3.5 meV/kbar, while the higher energy transition (transition B, shown as circles) exhibits a lower pressure coefficient, around −1.3 meV/kbar. The discrepancy in the energy of transition B between both samples is accounted for by uncertainties in the fitting procedure of the weaker transition and sample misorientations.

### b. Polarization measurements

In order to evaluate all the excitonic transitions present in the photoreflectance spectra, polarization-resolved measurements were performed at selected pressures (spectra shown in the supplementary material, Figs. S3, S4). Following the same fitting procedure described for $ReS_2$ above, we found that two transitions are enough to describe all the features present in the PR spectra at all polarization angles and all the studied pressure range. This is in good agreement with previous[65] polarization-resolved measurements performed at ambient pressure, which also found two excitonic transitions. The polarization dependence of the amplitudes for each transition is shown in Fig. 9 for the pressures 4.5 kbar and 8.5 kbar (upper and lower half of the chart, respectively). As it can be seen in the figure, the orientation and relative amplitude for each transition is preserved at different pressures, which is expected since $ReSe_2$ does not undergo any phase transition up to ≈100 kbar.[30] After fitting Eq. [2] (plotted as solid curves) we found that both transitions are strongly polarized ($f_\perp \approx 0$) along the angles $\varphi_0 \approx 98°$ and 175°, hence with a relative angle of 77°. This result evidences that the orientation of the $b$ axis for the sample III is $\varphi_0 \approx 98°$ since previous polarization-resolved measurements found that $E^{ex}_1$ is polarized along the $b$ axis.[25,65] Using the same procedure for the sample IV we found that its orientation is $\varphi_0 \approx 157°$, which is significantly different from that of sample III. The different sample orientations partly account for the differences in the fitted energy for the transition B (see Fig. 8) between both samples.

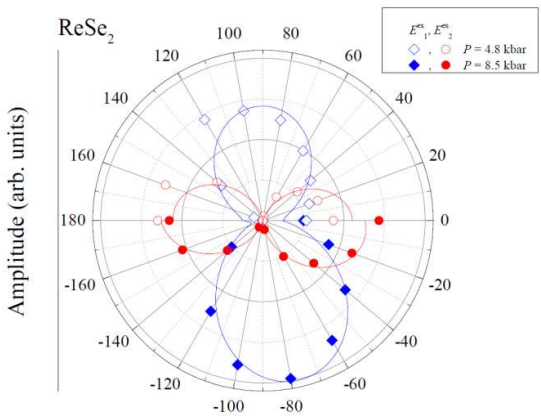

*Fig. 9.* Polar dependence of the amplitude for two fitted PR transitions in Sample III. The fittings have been performed at two different pressures, 4.5 kbar (empty symbols, top) and 8.5 kbar (full circles, bottom). The solid curves are fits from Eq. [2].

### c. First-principle calculations

The calculated (experimental[66]) lattice parameters for ReSe$_2$ at 0 kbar calculated with SCAN+rVV10 are as follows: $a$ = 6.578 Å (6.597 Å), $b$ = 6.749 Å (6.710 Å), $c$ = 6.784 Å (6.721 Å) and $\alpha$ = 95.30° (91.84°), $\beta$ = 103.30° (104.9°), $\gamma$ = 119.74° (118.91°). Under hydrostatic compression the $c$ lattice parameter decreases most significantly, down to 97.4 % of its original value at a pressure of 20 kbar. In contrast, lattice parameters $a$ and $b$ are compressed by less than 0.5 % at 20 kbar. The angles display only slight variations (see Section 3 of the S.I. for more details).

Our high-density k-mesh calculations predict an indirect fundamental band gap of 1.10 eV for ReSe$_2$ at 0 kbar with both the VBM and the CBM located away from high symmetry points (named J2 and J3, respectively, see Table 1). This is in agreement with recent studies considering the whole 3D Brillouin zone, which also found an indirect band gap with the VBM outside the symmetry points.[27,64] While the exact coordinates of the VBM are not mentioned in these studies, they are qualitatively close to the J2 k-point reported in the present work. Another study found two energetically degenerate VBM from ARPES measurements, one at Z and another off the symmetry points.[37] The same authors conducted theoretical calculations of the band structure and found that the nature of the band gap depends on the functional employed (indirect for GGA, direct for GW). Several other studies that take only high-symmetry k-paths into account for evaluating the band structure predict either direct or indirect band gaps for ReSe$_2$ near the Z or Γ point.[24,29,34,67,68] It has also been suggested that the indirect and the direct band gap are close in energy, and the discussion about the nature of the fundamental band gap for ReSe$_2$ is still ongoing.[64,68,69,67] Thus, a careful analysis of the 3D Brillouin zone is critical before making a decision on the nature of the band gap. Our calculations reveal that the lowest-energy direct band gap is located at a k-point off the symmetry lines (J1: 0.02941, 0.14706, −0.20588, see Table 1.) with a band gap energy of 1.15 eV. The second-lowest direct band gap of 1.19 eV is located at the Z point. These values are lower than the experimentally measured optical gap, of 1.31 eV and 1.36 eV. Note that, since the exciton binding energy is $E_b \approx$ 120 meV,[24] the experimental fundamental gaps are $E_g$ = 1.43 eV and 1.47 eV, as obtained from $E_g = E_{opt} + E_b$. Hence, the calculated fundamental band gap is ≈20% lower than the experimental. The discrepancies between calculations and experiments are accounted for by the band gap underestimation of the SCAN functional.[46] We also performed calculations at a higher level of theory using the hybrid functional HSE06[62]. At this level of theory, our calculations are able to predict the band gap energies more accurately with 1.55 eV and 1.59 eV at Z and K1 (including the excitonic binding energy) (see Table S3 of the S.I.).

The electronic dispersion curves for ReSe$_2$ are shown in *Figure 10* together with the optical matrix elements and bandgaps along the 3-D Brillouin zone and k-points of interest. From the figure it can be seen that an overall narrowing of the band gap takes place with increasing pressure. The matrix element maxima correspond with the band gap minima.

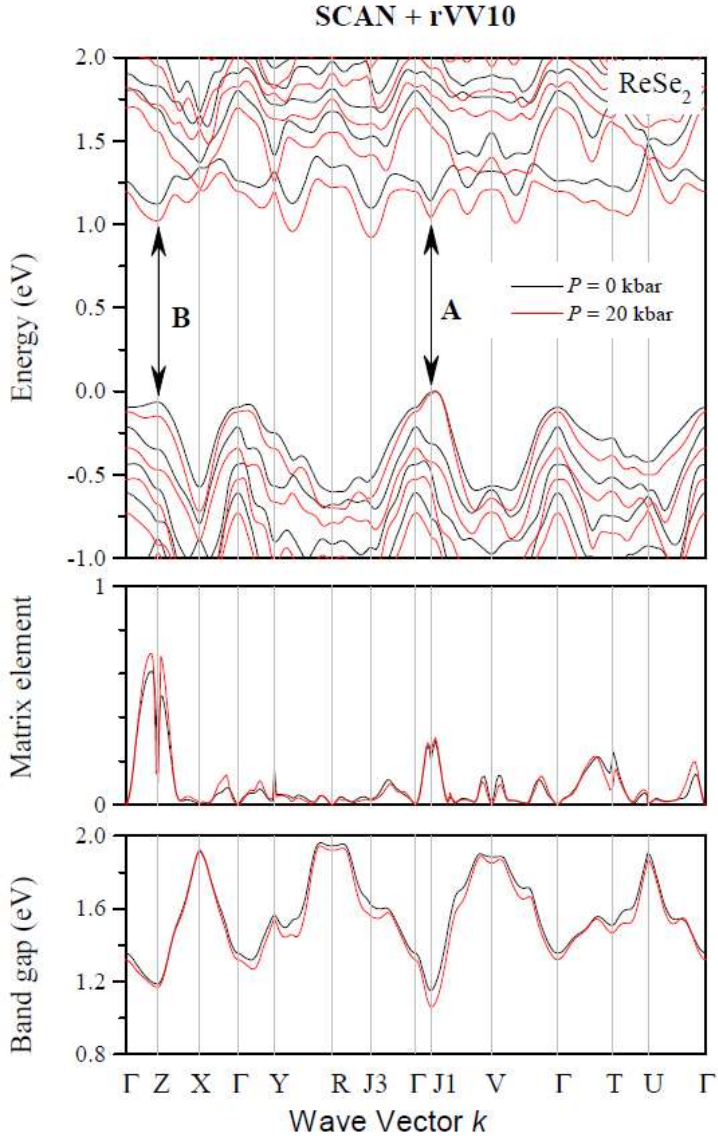

*Fig. 10.*

Electronic dispersion curves for ReSe$_2$ as calculated using the SCAN functional at zero pressure (black curves) and 20 kbar (red curves). The corresponding matrix elements have been calculated for each *k*-point, the stronger transitions are located around the Z and J1 *k*-points. In the lower panels, the direct band gap energy is plotted along the studied wave vectors.

In order to assign the experimentally measured transitions A and B, pressure coefficients were calculated for the band gaps at J1 and Z at different pressures. Defining the pressure coefficient as the slope of the linear fit of the band gap energies at 0, 5, 10, 15 and 20 kbar, we obtained calculated pressure coefficients of −4.2 meV/kbar$^{-1}$ and −0.7 meV/kbar$^{-1}$ for J1 and Z, respectively. The values for the fit were chosen according to the maximum in the matrix element as described for ReS$_2$ above. Statistical errors for the calculated pressure coefficient were relatively small for ReSe$_2$ with about ±0.3 meV/kbar for J1 and ±0.4 meV/kbar for Z. The calculated pressure coefficients are similar to the experimental ones, −3.51 meV/kbar and −1.35 meV/kbar for the transitions A and B, respectively. The pressure dependence of the measured (full symbols) and calculated (open symbol) variation of energy for each transition is shown in the *Figure 11*. As it can be seen in the figure, the energy of transition A (full square symbols) exhibits a stronger redshift with increasing pressure than that of transition B (full open symbols). Note that the pressure dependence of each transition is very similar between samples (blue and red symbols for samples III and IV, respectively). Such splitting is in good agreement with the calculated values, which allows to unambiguously assign the transition A to the J1 k-point of the Brillouin zone and transition B to the Z k-point. Our HSE06 calculation confirm this trend (see Table S3 of the S.I.).

The current assignation of the transition A at the J1 k-point is in contrast with previous assumptions that all the excitonic transitions took place around the Z point of the Brillouin zone.[24] The present result should be taken into account for future work on the compositional dependence of the band gap of the ReSe$_{2-x}$S$_x$ alloy since the J1 k-point is away from either the Z or K1 k-points (see Table 1). Previous absorption[61] and piezoreflectance[35] measurements along the entire composition range found evidence that the nature of the band gaps are similar for the ReS$_2$ and ReSe$_2$ compositional end members. However, while we found that both direct excitonic transitions are similar in energy (the transition energy in the J1 k-point is only ≈40 meV below that at Z) they belong to different k-points between different compositional end members. Hence, it is expected that the compositional dependence of the lowest direct transition (i.e. transition A) exhibits a crossover from the J1 k-point for ReSe$_2$ to the Z k-point for ReS$_2$.

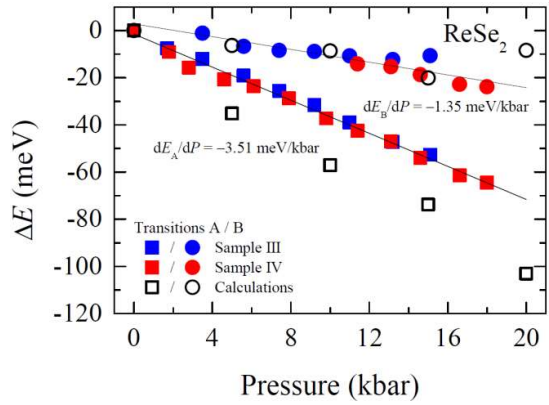

*Fig. 11.*
Variation of exciton energy versus pressure plotted for the transitions A (squares) and B (circles) from measurements on samples III (blue) and IV (red) as well as the calculated values using the SCAN functional (open symbols). The straight lines are linear fits to the experimental values, the pressure coefficients of each transition is included.

## V. DISCUSSION

Fig. 12 shows the pressure coefficient of the first direct optical transition for $MX_2$ TMDCs (X = S, Se and M = Mo, W, Re) as measured by high-pressure photoreflectance spectroscopy elsewhere,[38] together with our experimental results for $ReX_2$. While $MoX_2$ and $WX_2$ exhibit positive pressure coefficients, this is not the case for $ReX_2$, which exhibit negative pressure coefficients. Such striking difference is accounted for by the particular crystallographic structure of $ReX_2$, and the particular electronic configuration for Re: With respect to group 6 transition metals, rhenium compounds possess one more valence electron and the valence and conduction band states are importantly characterized by the Re-d orbitals. In order to investigate the physical origin of the negative pressure coefficient, and its relation with the reduced van der Waals interaction in $ReX_2$, we performed an orbital analysis of the states associated with the A and B transitions.

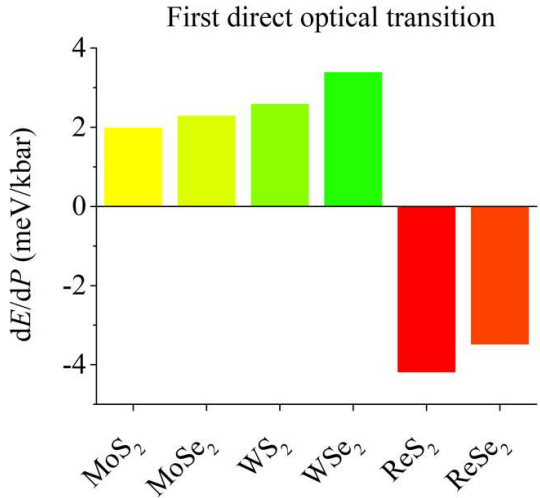

*Fig. 12.* Histogram showing the pressure coefficient of the first direct optical transitions of $MoX_2$ and $WX_2$ published elsewhere[38] and $ReX_2$ (X = S, Se) as obtained from high-pressure photoreflectance measurements.

The orbital composition of the states of the A and B direct transitions is shown in Table 2 for $ReS_2$ and $ReSe_2$. The conduction band minimum and valence band maximum of $ReS_2$ are dominated by Re-$d_{z^2}$ orbitals, in agreement with recent calculations,[26] while for $ReSe_2$ the orbital contributions are more diverse. In Table 2 the Re-$d_{z^2}$ and X-$p_z$ orbital contributions, which importantly contribute to the band edge states[70] and show out-of-plane character,[69,71] are highlighted since these are expected to be highly sensitive to interlayer interaction. With increasing pressure, the states with large contributions from Re-$d_{z^2}$ and X-$p_z$ orbitals destabilize, and therefore rise in energy with increasing pressure. Such destabilization has been attributed to Coulomb repulsion of anti-bonding p-orbitals between interlayer chalcogen atoms for $MoS_2$.[72–74] Similarly, $d_{z^2}$ orbitals are fairly delocalized and directed in the perpendicular direction. The role of orbital contribution on the band gap dependency on interlayer distance is well studied for other TMDCs, and is the state-of-the-art explanation for the direct-to-indirect band gap crossover of $MoS_2$ at its transition from monolayer to bulk.[72,75] As a rough approximation, increasing pressure has a similar effect on the electronic structure of $ReX_2$ as increasing the number of layers on conventional TMDCs, i.e. pressure results in a stronger interaction of electrons along the van der Waals gap and a reduction of interlayer distance. In fact, the pressure and strain dependence of the bandgap of $MoX_2$ has been explained in terms of orbital contributions to the bandgap states.[76,77] Hence, larger contributions of the Re-$d_{z^2}$ and X-$p_z$ orbitals in the VBM with respect the CBM would result in a narrowing of the band gap with increasing pressure, which is the case for $ReX_2$ as discussed in detail below.

The orbital interplay on the band gap reduction of $ReX_2$ with increasing pressure/strain has been previously hinted[12,33] but never evaluated from orbital analysis. For the case of $ReS_2$, the highest contribution to the analyzed states arises from Re-$d_{z^2}$. In Table 2 it can be seen that larger contributions of the $p_z$ and $d_{z^2}$ orbitals take place in the VBM with respect the CBM. Hence, at higher pressures the VBM should experience a stronger destabilization than the CBM, resulting in a narrowing of the band gap, as observed experimentally. Furthermore, the transition at K1 (i.e. transition B) exhibits significantly more contribution from the S-$p_z$ states, which accounts for its more negative pressure coefficient with respect the transition at Z (i.e. transition A), as observed experimentally (see Fig. 6). Similarly, for the case of ReSe$_2$, large contributions from the Se-$p_z$ and Re-$d_{z^2}$ orbitals take place in the VBM, which implies a large redshift of the transition at J1 (i.e. transition A) with increasing pressure, in agreement with the experimentally observed large negative pressure coefficients for the transition A (see Fig. 11). In contrast, the transition at Z (i.e. transition B) shows only moderate contributions of Se-$p_z$. In order to account for the negative pressure coefficient of transition B we conclude that the Re-$d_{yz}$ orbitals play a significant role since the CBM exhibits larger contributions from the Re-$d_{z^2}$ orbitals than the VBM. In general, the band gap narrowing with increasing pressure on the transitions A and B of ReX$_2$ is mainly accounted by an increased contribution of S-$p_z$ orbitals in the VBM.

So far it has been shown that the negative pressure coefficients observed for the direct transitions in ReX$_2$ can be qualitatively explained from orbital theory. However, the value of the pressure coefficient could be influenced by the reduced van der Waals interactions present in ReX$_2$ with respect other TMDCs. In this regard, high-pressure measurements are highly valuable because they allow to directly modulate the interlayer interaction. For the case of ReX$_2$, while it is broadly agreed that it exhibits decreased interlayer coupling with respect group 6 TMDCs, the degree of such interlayer coupling is nowadays controversial, and available data must be interpreted carefully.

Recent works show, on the one hand, strong evidences of reduced van der Waals interactions. For instance, photoluminesce experiments that showed that the emission energy of ReS$_2$ is almost independent to the number of layers ($\Delta E \approx -50$ meV from 1 monolayer to bulk) in contrast with other G6-TMDCs (e.g. $\Delta E \approx -600$ meV for MoS$_2$),[3] which suggests that layers can be electronically decoupled. Moreover, opposite to other TMDCs, the PL signal of ReS$_2$ increases with the number of layers which could be either attributed to a strong layer decoupling effect or an almost direct bandgap in bulk ReS$_2$. Similarly, it was reported that ReSe$_2$ retains a direct band gap regardless of its crystal thickness, with excitons strongly confined within single layers for bulk crystals (68%).[24] Besides, Raman measurements showed that the Raman spectrum of monolayer ReS$_2$ is almost identical to that of bulk and lattice dynamical calculations evidenced ultraweak interlayer coupling.[28] High-pressure works on ReS$_2$ also reinforce the decoupled behavior in bulk ReS$_2$. High-pressure Raman measurements showed a twofold lower phonon pressure coefficient for the out-of-plane $A_{1g}$ mode with respect other TMDCs, reinforcing the decoupled behavior in bulk ReS$_2$.[3] The large pressure metallization of ReS$_2$ (70 GPa) in comparison to MoS$_2$ (19 GPa) has been attributed to the larger interlayer coupling in MoS$_2$, since the interlayer overlap of wave functions of MoS$_2$ is more prominent at ambient pressures and therefore metallization can be easier attained under high pressures.[33] One of the most direct ways to probe the van der Waals interlayer forces is by measuring the bulk modulus. For the case of ReS$_2$ and ReSe$_2$, bulk modulus of 23 GPa[31] and 31 GPa[30] were measured by high-pressure X-ray diffraction measurements, respectively. These figures are significantly reduced in comparison to other TMDCs such as MoS$_2$ (57 GPa[78]), MoSe$_2$ (62 GPa[79]), WS$_2$ (63 GPa[80]) or WSe$_2$ (72 GPa[81]). On the other hand, photoemission experiments[23,69,70] showed that there exist a significant electronic dispersion along the van der Waals gap and conclude that layers in ReX$_2$ are not completely electronically decoupled. Also, recent calculations showed that the fundamental band gap of ReS$_2$ (ReSe$_2$) strongly depends on the number of layers, which shifts from 2.38 eV (2.05 eV) for monolayers down to 1.60 eV (1.38 eV) for bulk, and remains direct.[29] Hence, despite it seems clear that while ReX$_2$ exhibit somewhat structural and vibrational decoupled behavior, the electronic coupling is still significant.

In order to elucidate whether ReX$_2$ exhibits a strong electronic decoupling, we present a comparative study of the orbital interplay on the pressure coefficient on ReX$_2$ and other group 6 TMDCs. As shown in Fig. 12, the pressure coefficient of the first direct optical transition in MX$_2$ (M = Mo, W, X = S, Se) is positive. This transition takes place at the K-point of the BZ of their 2H crystal structure.[38] In the case of MoS$_2$, the orbital contributions at this k-point are 79 % Mo-$d_{x^2-y^2}$ + 21 % S-$p_{xy}$ for the VBM and 86 % Mo-$d_{z^2}$ + 9 % S-$p_{xy}$ + 5 % S-$p_z$ for the CBM.[72] Hence, the positive pressure coefficient in MoS$_2$ is mostly accounted for by an increase of the CBM energy with pressure. However, at high pressure a metallization takes place from the increase of the VBM at Γ where the orbital contributions from $d_{z^2}$ and $p_z$ orbitals are strong, i.e. 60% Mo-$d_{z^2}$ + 30% S-$p_z$. The same behavior is expected for MX$_2$ (M = Mo, W, X = S, Se), where the VBM at Γ is expected to exhibit strong orbital contributions from metal $d_{z^2}$ and chalcogen $p_z$ orbitals. This is supported by the DFT calculations performed elsewhere,[38] which reported pressure coefficient of the indirect band in the range −3.79 to −7.9 meV/kbar. These values are equal or higher than the here reported pressure coefficients (i.e. −4.2 meV/kbar and −3.5 meV/kbar for ReS$_2$ and ReSe$_2$, respectively). From this result we argue that the decreased van der Waals interaction in ReX$_2$ does not play a significant role in its pressure coefficient, since an electronic decoupling would yield significantly larger pressure coefficients in contrast to other TMDCs due to either stronger reduction of quantum confinement[75,82] or increased Coulomb repulsion of the antibonding chalcogen orbitals[72–74] with decreasing interlayer distance.

*Table 2.* Calculated orbital composition of the important extrema (transitions A and B) of the electronic band structure of ReS$_2$ and ReSe$_2$. Highly interlayer-affected orbitals ($p_z$, $d_{yz}$ and $d_{z^2}$) are highlighted in blue. The $d$ orbital contributions come solely from Re atoms, while $p_z$ orbitals are primarily of chalcogen atoms.

| Material | k-point | Assigned transition | Orbital composition |
|---|---|---|---|
| ReS$_2$ | Z | A | VBM: 33 % $d_{z^2}$ + 20 % $d_{xz}$ + 15 % $p_z$ + 14 % $d_{xy}$ + … |
| | | | CBM: 33 % $d_{z^2}$ + 22 % $p_x$ + 18 % $d_{x^2-y^2}$ + 14 % $d_{xy}$ + … |
| | K1 | B | VBM: 32 % $d_{z^2}$ + 22 % $p_z$ + 15 % $d_{xz}$ + 12 % $d_{xy}$ + … |
| | | | CBM: 31 % $d_{z^2}$ + 20 % $p_x$ + 19 % $d_{x^2-y^2}$ + 14 % $d_{xy}$ + … |
| ReSe$_2$ | J1 | A | VBM: 50 % $p_z$ + 15 % $d_{z^2}$ + 12 % $p_y$ + 7 % $d_{x^2-y^2}$ + … |
| | | | CBM: 21 % $d_{x^2-y^2}$ + 21 % $d_{z^2}$ + 16 % $p_y$ + 12 % $d_{yz}$ + 11 % $d_{xz}$ + … |
| | Z | B | VBM: 31 % $d_{yz}$ + 25 % $d_{x^2-y^2}$ + 17 % $d_{z^2}$ + 9 % $p_y$ + 9 % $p_z$ + … |
| | | | CBM: 32 % $d_{z^2}$ + 30 % $p_y$ + 14 % $d_{x^2-y^2}$ + 9 $d_{xy}$ + 6 % $p_z$ + … |

## VI. SUMMARY AND CONCLUSIONS

We performed high-pressure photoreflectance measurements on ReS$_2$ and ReSe$_2$ samples obtained from different sources and grown on different conditions. Our results reveal that two main excitonic transitions decrease in energy with increasing pressure for each material. For the case of ReS$_2$, the obtained pressure coefficients for the A and B transitions are −2.3 meV/kbar and −4.2 meV/kbar, respectively and for ReSe$_2$, −3.5 meV/kbar and −1.3 meV/kbar, respectively. Polarization-resolved measurements allowed to measure a third transition for ReS$_2$, as well as determining the crystal orientation and assess the structural stability up to 20 kbar in ReX$_2$.

The electronic band structure of ReS$_2$ and ReSe$_2$ was calculated from *ab initio* calculations within the density functional theory, using the meta-GGA SCAN functional. We probed the whole Brillouin zone in order to explore all the possible direct transitions around the band gap. The calculations were performed at different pressure values, which allowed the comparison of the experimental and theoretical results and assignment of each transition. For ReS$_2$, the transitions A and B were assigned to Z and K1 k-points of the BZ whereas for ReSe$_2$ the A and B transitions were assigned to the J1 and Z points, respectively (with K1 and J1 both located away from the high-symmetry k-points). The negative pressure coefficients measured in ReX$_2$ were explained in terms of orbital analysis. This allowed us to conclude that the destabilization of the $p_z$ orbital with increasing pressure are mostly responsible for the measured pressure coefficients. The present work evidences that ReX$_2$ does not exhibit a strong electronic decoupling, in agreement with recent angle-resolved photoemission studies.

## Acknowledgements

Work supported by the National Science Centre (NCN) Poland OPUS 11 no. 2016/21/B/ST3/00482. R.O. acknowledges the support by POLONEZ 3 no. 2016/23/P/ST3/04278. This project is carried out under POLONEZ programme which has received funding from the European Union's Horizon 2020 research and innovation programme under the Marie Sklodowska-Curie grant agreement No 665778. F.D. acknowledges the support from NCN under Fuga 3 grant no. 2014/12/S/ST3/00313. We thank Xavier Rocquefelte for discussions regarding the structure of transition metal dichalcogenides. M. L. and O. R. would like to acknowledge the funding provided by the Natural Sciences and Engineering Research Council of Canada under the Discovery Grant Program RGPIN-2015-04518. The computations were performed using Compute Canada (Calcul Quebec and Compute Ontario) resources, including the infrastructure funded by the Canada Foundation for Innovation. Finally, S. T. acknowledges funding provided by National Science Foundation DMR-1552220 and DMR-1838443.

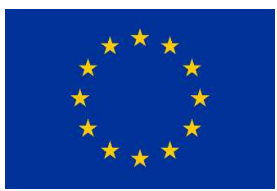

## Author contributions

R.O. wrote the manuscript, took part in PR measurements and analyzed PR data, M.L. carried out first-principle calculations and contributed with the drafting of the discussion, F.D. and J.K. performed the high-pressure PR experiments, Y.Q. and S.T. grow the samples, O.R. planed and supervised the calculations, R.K. planned the research and coordinated it. All authors discussed the results and commented on the manuscript.

## Additional Information

**Competing financial interests:** The authors declare no competing financial interests